\documentclass[twocolumn,aps,showpacs,pra,preprintnumbers]{revtex4-1}
\usepackage{graphicx}
\usepackage{amsmath}
\usepackage[utf8]{inputenc}

\usepackage{xcolor}     
\usepackage{soul}       

\begin{document}

\title{Scaling properties for a harmonic trapped gas near Bose-Einstein condensation using thermal global expansion coefficient measurement}

\author{E. D. Mercado-Gutiérrez}
\author{F. J. Poveda-Cuevas}
\author{V. S. Bagnato}
\affiliation{Instituto de Física de São Carlos, Universidade de São
Paulo, C.P. 369, 13560-970 São Carlos, SP, Brazil}

\date{\today}

\begin{abstract}
We report the measurement of the global thermal expansion coefficient of a confined Bose gas of $^{87}{\rm Rb}$ in a harmonic potential around the Bose-Einstein condensation transition temperature. We use the concept of global thermodynamic variable, previously introduced and appropriated for a non homogeneous system. The data show divergence on the thermal expansion coefficient near the transition temperature and analyses of the universality behavior shows a critical exponent $\alpha \sim 0.15\pm 0.09$. With this obtained value and the critical exponent for the correlation length recently presented in the literature \cite{Donner16032007} we analyse the scaling law combining dimension and critical exponents. 
\end{abstract}

\pacs{03-75.Hh, 67.85.Hj}

\maketitle

\section{Introduction}
The study of phase transitions and the behaviour of the properties of physical systems near and across phase transitions are of fundamental interest in physics. In general, near a phase transition, fluctuations become dominant producing, as a consequence, variations on the macroscopic properties, including the  thermodynamic susceptibilities \cite{Baym1999}. In this sense, the study of thermodynamic observables such as heat capacity, compressibility or even thermal expansion are quite important in the build up of a comprehensive picture of the phase transitions and their related properties. 

It is well known, for example, that, near a second order phase transition, suceptibilities will present an universal behavior scaling with the reduced temperature $t_{r}=|(T-T_c)|/T_c$, with an exponent depending only on the dimensionality \cite{JzinnQFT}. In these cases, the properties of the system near the critical points depend only upon general characteristics of the system like its dimensionality and number of degrees of freedom, rather than its detailed structure and/or components. This universality allows one to join up the phase transitions in classes which, in its turn, allows to study a whole group of phenomena by studying the phase transition properties of only one of them, since they share the same universality class. As an example of the outreach of the above, we have Bose-Einstein condensation (BEC) in 3D belonging to the same universality class as many magnetic phase transition \cite{Stanley1999}. 

To date, investigation of critical exponents has been essentially restricted to homogeneous systems. $^{4}{\rm He}$ around the $\lambda$-point constitute the mostly explored system with respect to critically. Recently, trapped atomic gases have taken the stage in this class of investigations, mainly due to the versatility of the systems in what concerns dimensionality and tunability of important experimental parameters. In fact, trapped gases in non-homogeneous potentials are quite intriguing because, as such, the density distribution is not homogeneous, opening up a whole set of different conditions to investigations concerning critical behavior \cite{Goswami2013}. 

In a recent paper by Donner et al \cite{Donner16032007}  the BEC was investigated in the critical regime observing the correlation length of critical fluctuation as a function of temperature. The observation enable them to obtain a critical exponent through the correlation length, revealing a critical exponent ($\xi\sim t^{-\nu}$) of $\nu=0.67\pm 0.13$.  Also, heat capacity and thermal expansion are other explored properties for trapped atoms. While Local Density Approximation (LDA) has been used \cite{Ku563} to investigate such thermodynamic properties by reducing a non-homogenous system to a large set of homogeneous subsystems each one with its own properties, there are intrinsic difficulties with respect to the determination of isobaric properties, since using LDA, the system experience a large variety of pressures within the same sample along its spatial extension. 

Isobaric properties are of special interest, since, for example, if one can determinate the critical exponent for the isobaric heat capacity $C_p \sim t^{-\alpha}$, it is possible to test the hyperscaling law $d\nu = 2-\alpha$ (where $d$ is the dimensionality and $\alpha$ the critical exponent). In fact, to measure $\alpha$, one can investigate the isobaric thermal expansion coefficient which also provides $\alpha$ \cite{Ahlers1976}. Indeed, thermal expansion coefficient in the context of scaling and universalities is being a topic of great interest in superfluid liquid $^{4}{\rm He}$ for quite a long time \cite{Mueller1976} and comparison with theory has been done mainly using re-normalization group theory of critical phenomena \cite{Wilson1971, Fisher1974}. The field is still quite exciting and new measurements keep being performed as recently through heat capacity determination in a zero gravity environment for Helium \cite{PhysRevB.68.174518}.

\section{Global thermal expansion measurement}

Here we report the measurement of the thermal expansion coefficient $\mathbf{\beta}_p = \frac{1}{V}\left(\frac{dV}{dT}\right)_{p,N}$ as a function of temperature using a global thermodynamic variable approach \cite{Vander2005}. Based on its behaviour near to $T_c$ we determine the critical exponent considering the dependence with the reduced temperature $t_r$. This approach of global variables has been used recently for experimental investigations using a sample of $^{87}{\rm Rb}$ atoms confined in a magnetic trap to determinate the state equation\cite{vander2012}, the heat Capacity at ``constant volume" \cite{PhysRevA.90.043640} and the isothermal compressibility \cite{EmmanuelPRA2015} during the occurrence of BEC in the harmonically trapped sample. 

In brief, the global variable description for the thermodynamics of a trapped system is quite convenient for experimental work since it avoids the strong local fluctuations present in the use of local-density approximation (LDA)\cite{Ku563}. The whole idea is to identify a set of macroscopic (Global) variables to describe the thermodynamics functions of the system. For the case of a harmonic trapped gas, the extensive variable equivalent to the volume in homogeneous system is the defined ``volume parameter",
\begin{equation}
{\cal V} =\frac{1}{\bar{\omega}^3};
\end{equation}

 with $\bar{\omega}^3=\omega_x\omega_y\omega_z$, (all frequencies of the trap potential).
In this case the thermodynamic limit is obtained taking $N\rightarrow\infty$; ${\cal V}\rightarrow\infty$ such that $N/\cal{ V}$ keeps constant. With a set of thermodynamical variables as $(N, T,{\cal V})$ a new pressure parameter ${\rm\Pi}(N,T,{\cal V})$ is defined such that ${\rm\Pi}=-\left( \frac{\partial F}{\partial{\cal V}}\right)_{N,T}$ where $F$ is the Helmholtz free energy. For the harmonic trap the pressure parameter is defined by \cite{vander2012, PhysRevA.90.043640, EmmanuelPRA2015}
\begin{equation}
\Pi=\frac{2}{3{\cal V}}\int d^3r\, n(\vec{r})\frac{1}{2}m\,(\vec{\bar{\omega}}\cdot\vec{r})^2,
\end{equation}
With $n(\vec{r})$ being the number density distribution. In this approach $\Pi$ and $\cal V$ correspond to a pair of conjugated variables (intensive and extensive quantities).

In terms of global variables the thermal expansion coefficient becomes
\begin{equation}
\beta=\frac{1}{{\cal V}}\left(\frac{\partial{\cal V}}{\partial T}\right)_{N,\Pi}.
\end{equation}
To determine $\beta$ we must obtain the isobaric $\cal V$ vs $T$ curves and from those, calculate $\beta$ for different temperatures for each pressure, with a fixed number of particles in the system. 

The investigation of $\beta$ as a function of the temperature across the BEC transition, is done using an experimental setup already described in detail in previous publications \cite{EmmanuelPRA2015}. Basically a double $^{87}\rm Rb$ MOT system allows to capture in a first stage $10^8$ atoms which are transferred to a second MOT with an appropriate vacuum of $10^{-11}~\rm{Torr}$. After sub-Doppler cooling the sample reaches around 35 $\rm \mu K$ and is transferred to a hybrid magnetic-optical trap. The magnetic part of the trap is composed by a quadrupole while the optical part is produced by a single focused  $\lambda = 1064$ nm laser beam. Inside the hybrid trap atoms are cooled by forced evaporation through a ramping down of the laser power, where the BEC is achieved. The hybrid trap constitutes  a non-isotropic harmonic oscillator with frequencies $(\omega_x, \omega_y,\omega_z)$. Those frequencies depend on the laser power (considering fixed focusing geometry) and the gradient of the magnetic field. The trap has a cylindrical symmetry where the radial frequency $(\omega_x, \omega_z)$ depends on the optical dipole trap and the axial confinement $\omega_y$ depends on the magnetic field gradient. Changing the laser power we change the frequencies and consequently the volume parameter ${\cal V} = (\omega_x\omega_y \omega_z)^{-1}$.
 
The atomic cloud is characterized by optical absorption image after releasing from the trap and remain in free expansion for up to 30 ms. From the absorption image the bimodal density distribution allows determination of the condensate fraction and the temperature. We start by measuring the pressure parameter $\Pi$ for a large variety of temperatures ranging from 400 nK to 10 nK. Keeping the number of atoms as constant as possible around $1\times 10^5$. 

A set of volume parameters are measured for the data presented here. For each volume considered a whole procedure of optimization is performed. Ideally one should continuously vary the volume parameter but this is not technically  possible at the moment and therefore we are restricted to a small set of volumes. However, we consider six different set of frequencies generating six independent volume parameter ranging from $1\times 10^{-8}~\rm s^{3}$ to $18\times 10^{-8}~\rm s^{3}$. For each volume around $30$ values of temperatures are considered generating $\Pi$ vs $T$ and, from that dataset, the isobarics plots are obtained, considering constant pressures and generating a set of ${\cal V}$ vs $T$ points  for each pressure. We always consider a few points before and after the transition to be able to construct the isobaric plots containing both regimes: thermal and condensate.

\begin{figure}[h]
\includegraphics[width=1.0\columnwidth]{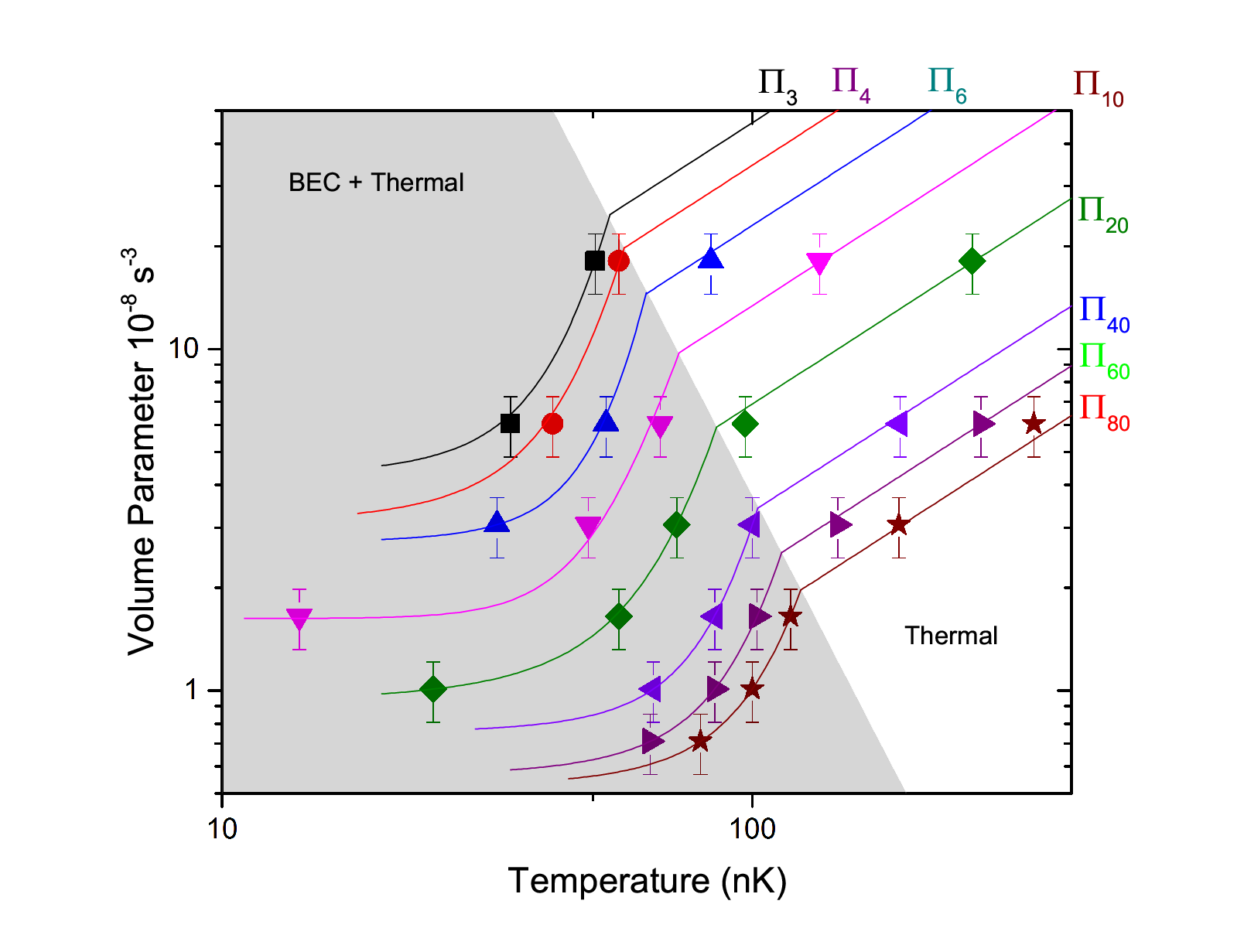}
\caption{Isobaric Curves for $\cal{V}$ vs T phase diagram, the atom number is $N=1\times 10^5$. Each curve represents a constant pressure ($\Pi$) transformation. There are two regions in the isobaric plot representing the existence of the two phases.}
\label{Fig 1}
\end{figure}

The obtained isobaric plots are presented in Fig. \ref{Fig 1}, where the experimental points were connected with a best curve (not a fitting), in order to produce a collection of points through the interpolation of the experimental data. From the isobaric plots, we graphically extract the derivative of the volume with respect to temperature for each chosen volume. Finally we obtain the thermal expansion coefficient $\beta$.

\section{Analysis of thermal expansion coefficient near the transition}

\begin{figure}[h]
\includegraphics[width=1.0\columnwidth]{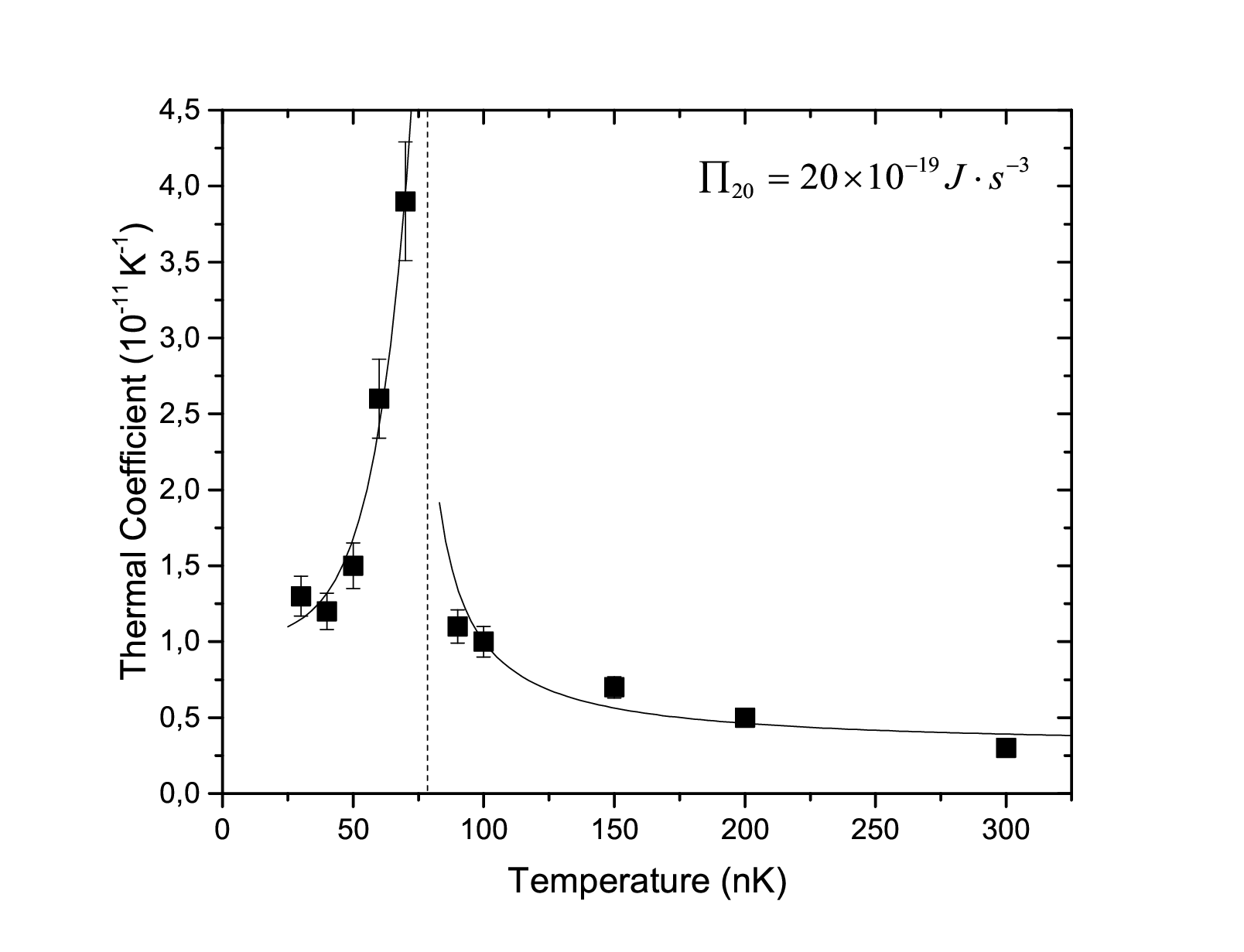}
\caption{Thermal expansion coefficient for pressure parameter $\Pi=20\times10^{-19}\,\rm J\cdot s^{-3}$. The divergence of $\beta$ near $T_c$ is the characterization of the phase transition. In this case $T_c\sim 80$ nK.}
\label{Fig 2}
\end{figure}

The procedure is repeated for many different temperatures resulting in a collection of data $\beta$ vs $T$ for each pressure parameter $\Pi$. A typical obtained plot is presented in Fig. \ref{Fig 2} for this case at pressure of $20\times10^{-19}$ $\rm J\cdot s^{-3}$

\begin{figure}[h]
\includegraphics[width=1.0\columnwidth]{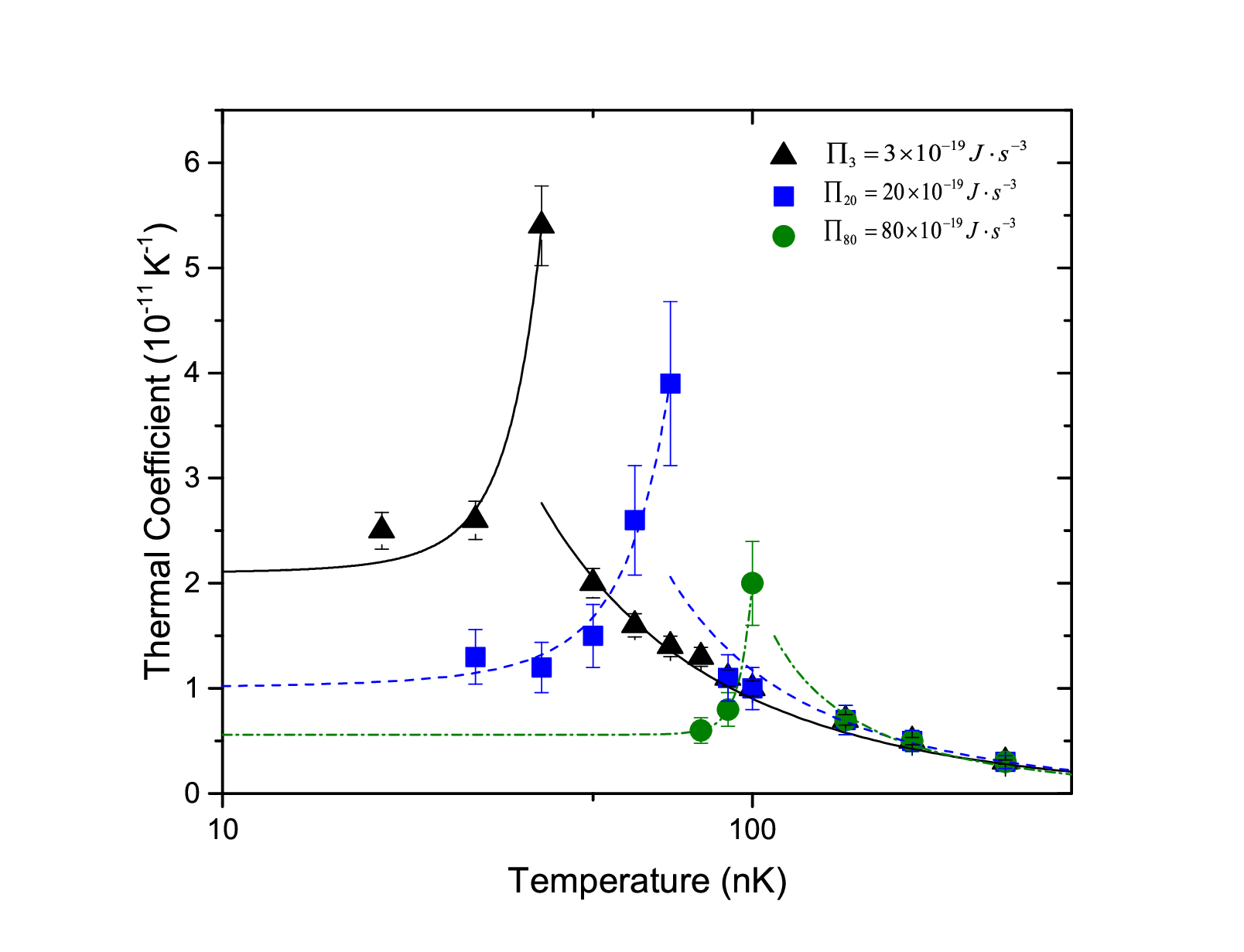}
\caption{Thermal expansion for different  pressures parameter. (The scale of each plot were displaced for better visualization)}
\label{Fig 3}
\end{figure}

For any of the considered pressures an equivalent plot is generate and presented for illustration all together in Fig. \ref{Fig 3}. The overall behaviour for $\beta$ across the condensation reminds a $\rm \lambda$-type behaviour with a  characteristic asymmetry and divergence at the critical temperature.

Looking at the isobaric curves (Fig. \ref{Fig 1}). the plots show two regions. At higher temperature above $T_c$ we have a linear dependence between ${\cal V}$ and $T$ indicating the thermal gas behaviour. At lower temperatures the change in the behaviour is consequence of the occurrence of the Bose-Einstein condensation. 

Interesting to note is that for a harmonic trapped gas, the occurrence  of BEC is followed by a compression of the global volume which means that to keep pressure (isobaric) constant as the temperature decreases, the volume parameter ${\cal V}=\bar{\omega}^{-3}$ has to decrease considerably during the formation of the BEC-phase. In this sense, on the analysis of the global variables, the condensation also represents a condensation in real space like a liquid droplet formed at the bottom of the potential.

The collection of the thermal coefficient plots in Fig. \ref{Fig 3} indicate that high pressure promotes condensation at high temperatures as expected. For all pressures, the typical shape in $\beta$ vs $T$ is present. To analyse the behaviour of $\beta$ with temperature we consider the case on Fig. \ref{Fig 2}, and analyse the behavior of $\beta$ around of $T_c$. The analysis of $\beta$ versus $t_r$ is done on the negative size ($T<T_c$). 

\begin{figure}[t!]
\includegraphics[width=1.0\columnwidth]{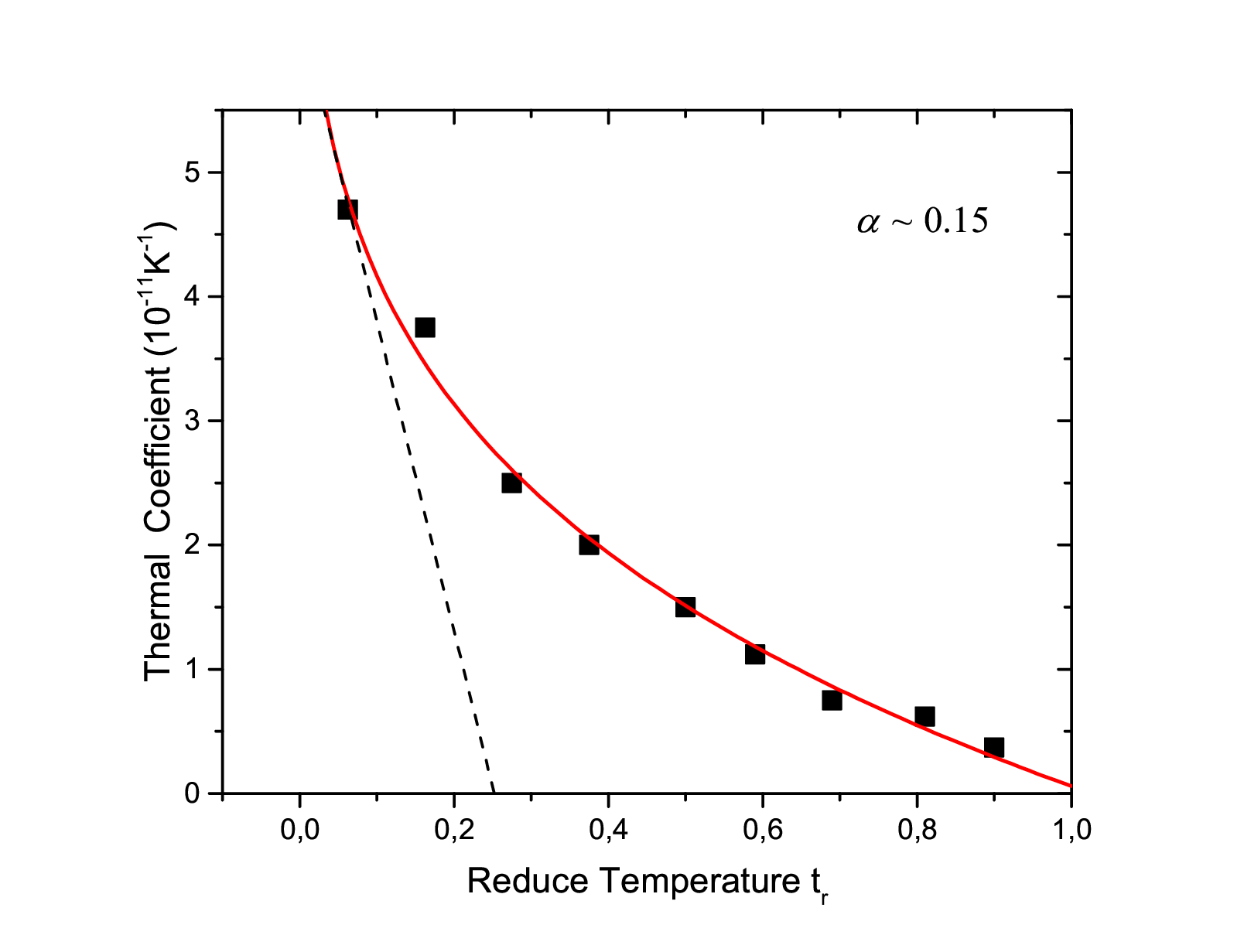}
\caption{Thermal expansion in function of reduced temperature. The asymptotic behavior at low values result in a exponential dependence $t^{-\alpha}$, with $\alpha\sim 0.15$.}
\label{Fig4}
\end{figure}

Considering that $\beta\sim t_{r}^{-\alpha}$ we determine
\begin{equation}
\alpha = -\lim_{t_r\rightarrow 0}\frac{d\ln\beta}{d\ln t_r},
\end{equation}
resulting in a value for $\alpha\sim 0.15$. In Fig. \ref{Fig4} we illustrate the behaviour of $\beta$ vs $t_r$, showing that in the limit $t_r \rightarrow 0$ the derivative is finite. Equivalent values for $\alpha$ are obtained for other pressure parameters, and the combined values result in $\alpha=0.15\pm 0.09$

Considering $\nu = 0.67\pm 0.13$ as measured in \cite{Donner16032007} and ours $\alpha = 0.15\pm 0.09$ we have from the law $d\nu = 2-\alpha$ a compatible dimension for the system
\begin{equation}
d=2.76\pm 0.16
\end{equation}
which is lightly smaller than 3, the expected value for our geometry. This deviation can come from the finite size of the system. Since the global variable is fully related to the thermodynamics in a formal sense, it should not introduce variations on $d$. We have also considered $\alpha$ from below the critical point, considering the scaling law that the critical exponent should be equal from both sides. In doing that another limitation arises from the technical difficulties to obtain data for $t_r$ very close to zero. Experiments for high precision determination of the critical exponent \cite{PhysRevLett.84.4894} are done in conditions where $t_r$ goes to values $10^{-3}$ to $10^{-4}$. We have reached $t_r\sim 0.06$ as the smallest value. The capability to measure $\beta$ for smaller $t_r$ could improve the value of $\alpha$. A deep theoretical evaluation on the critical exponent using global variables could also elucidate differences that are however out of the scope of this report.

Besides the technical limitation, it seems clear that the use of global thermodynamics variables may open up a whole new window of opportunities to investigate universality and scaling laws for those non-homogeneous systems. In special, interactions play an important role in this field. While it is practically impossible to vary the interaction in $^4\rm He$ $\lambda$-transition, it is possible for trapped atoms. The new possibilities for universality in strong correlated systems may be quite exciting since it is very unknown. In all new possible exciting situations, the global variable concept may well be a nice tool of great relevance.
\section{Conclusion}
As a conclusion, we have provided a measurement for the thermal expansion coefficient in cold atoms indicating new ways for studying the behavior of susceptibilities near the critical point. The expected scaling behavior when combining our measurements with existing critical exponent for correlation length \cite{Donner16032007} does not fully agree with the expected values, but there are many possibilities for improvement. The unique access to critical phenomena investigation with the approach here presented is the most relevant message opening up new possibilities in trapped cold atoms.\\

\section{Acknowledgements}

We wish to thank Fapesp, program CEPID and Capes for financial support. We greatly appreciate the collaboration with V. Romero-Roch\'{i}n (UNAM - M\'{e}xico), D. Varela Magalh\~{a}es, S. R. Muniz, K. Magalh\~{a}es, G. Telles, Emanuel Henn and G. Roati.
\bibliography{References_Thermal_Expansion_Coefficient_FV}
\end{document}